\documentclass[
 reprint,
 superscriptaddress,
 onecolumn,
 amsmath,amssymb,
 aps,
 prl,
 floatfix,
]{revtex4-2}

\usepackage{graphicx} 
\usepackage{dcolumn}  
\usepackage{bm}       
\usepackage{braket}   
\usepackage{xcolor}   
\usepackage{soul}
\usepackage{ulem}

\def\Cr135{CsCr$_3$Sb$_5$}
\def\V135{$A$V$_3$Sb$_5$}
\def\Ef{$E_\mathrm{F}$}

\renewcommand{\emph}[1]{\textit{#1}}
\definecolor{cgreen}{rgb}{0.0, 0.60, 0.32}

\begin{document}

\title{Flat-Band Enhanced Antiferromagnetic Fluctuations and Superconductivity in Pressurized \Cr135}

 \affiliation{School of Physics, Zhejiang University, Hangzhou 310058, P. R. China}
 \affiliation{Center for Correlated Matter, Zhejiang University, Hangzhou 310058, P. R. China}
 \affiliation{School of Physics, Hangzhou Normal University, Hangzhou 310036, P. R. China}
 \affiliation{Institute for Advanced Study in Physics, Zhejiang University, Hangzhou 310058, China}

\author{Siqi Wu}
 \affiliation{School of Physics, Zhejiang University, Hangzhou 310058, P. R. China}

\author{Chenchao Xu}
 \affiliation{School of Physics, Hangzhou Normal University, Hangzhou 310036, P. R. China}

\author{Xiaoqun Wang}
 \affiliation{School of Physics, Zhejiang University, Hangzhou 310058, P. R. China}

\author{Hai-Qing Lin}
 \affiliation{School of Physics, Zhejiang University, Hangzhou 310058, P. R. China}
 \affiliation{Institute for Advanced Study in Physics, Zhejiang University, Hangzhou 310058, China}

\author{Chao Cao}
 \email{ccao@zju.edu.cn}
 \affiliation{School of Physics, Zhejiang University, Hangzhou 310058, P. R. China}
 \affiliation{Center for Correlated Matter, Zhejiang University, Hangzhou 310058, P. R. China}

\author{Guang-Han Cao}
 \affiliation{School of Physics, Zhejiang University, Hangzhou 310058, P. R. China}

\date{\today}

\begin{abstract}
The spin dynamics and electronic orders of the kagome system at different filling levels stand as an intriguing subject in condensed matter physics. By first-principles calculations and random phase approximation analyses, we investigate the spin fluctuations and superconducting instabilities in kagome phase of \Cr135 under high pressure. At the filling level slightly below the kagome flat bands, our calculations reveal strong antiferromagnetic spin fluctuations in \Cr135, together with a leading $s_{\pm}$-wave and a competing ($d_{xy}$, $d_{x^2-y^2}$)-wave superconducting order. Unlike the general intuition that the flat bands are closely related to the ferromagnetic correlations, here we propose a sublattice-momentum-coupling-driven mechanism for the antiferromagnetic fluctuations enhanced from the unoccupied flat bands. The mechanism is generally applicable to kagome systems where the Fermi level intersects near the flat bands, offering a new perspective for future studies of geometrically frustrated systems.
\end{abstract}

\maketitle

\section*{Introduction}
The kagome system has attracted significant research interest due to its highly frustrated nature. In the Mott limit, several theoretical proposals has been made for the possible topological quantum spin liquid phases in spin-1/2 kagome Heisenberg model \cite{Ran-2007-QSL,Yan-2011-QSL,Xie-2014-QSL}. On the other hand, extensive investigations have also been carried out based on its unique electronic band structures away from the Mott limit. At the flat-band filling ($>$ 4/3 per site), the large degeneracy and nontrivial topology of the partially filled flat band could give rise to ferromagnetic (FM) ground state or fractional quantum Hall states \cite{Lieb-1989-FM,Mielke-1992-FM,Tang-2011-FQH,Parameswaran-2013-FQH}; at the van Hove filling (2/3 $\pm$ 1/6 per site), the perfect Fermi surface (FS) nesting dominates the instability of the system, yielding exotic phases such as bond orders, density wave (DW), as well as their intertwining with superconducting (SC) orders \cite{Yu-2012-SDW, Wang-2013-vHS, Kiesel-2013-vHS, Park-2021-vHS, Feng-2021-vHS, Jiang-2021-V135, Wang-2021-V135, Mielke-2022-V135, Denner-2021-V135, Guguchia-2023-V135, Wu-2021-V135, Jin-2022-V135}; around the Dirac filling ($\sim$ 2/3 per site), the system could develop topological nontrivial phenomena such as anomalous quantum Hall effect and topological insulating phases \cite{Guo-2009-TI, Xu-2015-QAH, Yin-2020-TbMn6Sn6, Ma-2021-RMn6Sn6, Xu-2022-QAH}. The interplay of electronic dynamics and geometry frustration in kagome system opens up new frontiers for uncovering a wealth of exotic physical phenomena.

Recently, a new kagome compound \Cr135\ was discovered \cite{Liu-2024-Cr135}. This compound features a structure analogous to that of the $A$V$_3$Sb$_5$ ($A = $ K, Rb, Cs) family \cite{Oritz-2019-V135, Wilson-2024-V135}, as shown in Fig.~\ref{fig:1}a. At ambient pressure, \Cr135\ exhibits a bad metal behavior and undergoes a DW like phase transition at around 55 K. The DW phase can be gradually suppressed at pressure larger than 3.65 GPa, accompanied by an SC dome as well as non-Fermi-liquid behavior in the phase diagram. Nuclear magnetic resonance measurements and DFT calculations show that the DW transition is likely to have a magnetic nature \cite{Liu-2024-Cr135,Xu-2023-Altermag}, suggesting the close relation between the spin dynamics and possible unconventional superconductivity (USC) in \Cr135. In contrast to the isostructural $A$V$_3$Sb$_5$ compounds ($A = $ K, Rb, Cs) whose low energy physics is dominated by the saddle-point patches around the Fermi level (\Ef) \cite{Park-2021-vHS}, \Cr135 is distinguished by the presence of incipient flat bands (IFBs). This presence has been supported by both experimental studies of angle-resolved photoemission spectroscopy measurements \cite{Li-2024-ARPES, Guo-2024-ARPES, Peng-2024-ARPES}, as well as theoretical studies such as density functional theory (DFT) calculations \cite{Liu-2024-Cr135, Xu-2023-Altermag}, dynamical mean-field theory \cite{Wang-2024-Kondo}, and slave-spin investigations \cite{Xie-2024-Cr135}. Previous theoretical studies in other systems have shown that the IFBs could substantially contribute to the virtual pair-scattering processes and play vital roles in the USC \cite{Aoki-2020-IFB,Kuroki-2005-IFB,Linscheid-2016-FeSe,Yamazaki-2020-BCO}. Moreover, the inter-band processes between non-Kramers degenerated bands could also contribute to nontrivial quantum geometry and lead to enhanced FM fluctuations and possible spin-triplet SC pairings \cite{Kitamura-2024-QM_Geometry}. Therefore, the presence of IFBs in \Cr135\ offers us a unique opportunity to study the effects of flat band physics on electronic correlations and superconductivity.

In this paper, we present our first-principles and random phase approximation (RPA) calculations for pressurized \Cr135. We begin by conducting RPA investigations on the Wannier-downfolded first-principles Hamiltonian at 5 GPa. To analyze superconducting (SC) instabilities, we employ a set of linearized SC gap equations. Our results indicate that the primary SC pairing channel in the pressurized \Cr135\ exhibits $s_{\pm}$-wave symmetry, accompanied by a competing ($d_{xy}$, $d_{x^2-y^2}$)-wave channel. Our theoretical analysis further highlights the essential role of the IFBs in AFM electron correlations. In the following discussions, we introduce a microscopic mechanism for the flat-band-enhanced antiferromagnetic (AFM) fluctuations in \Cr135, which incorporates the sublattice-momentum coupling (SMC) of both occupied dispersive bands and unoccupied IFBs. (In this paper, the electron susceptibilities are treated within the weak-coupling RPA scenario. The AFM spin fluctuations have an itinerant nature and should be understood as SDW modes.) This mechanism well explains the observed enhancement of AFM spin fluctuations in numerical calculations. It is also anticipated to be broadly applicable to kagome and other geometrically frustrated systems.

\section*{Results}
\begin{figure*}
  \includegraphics[width = 16 cm]{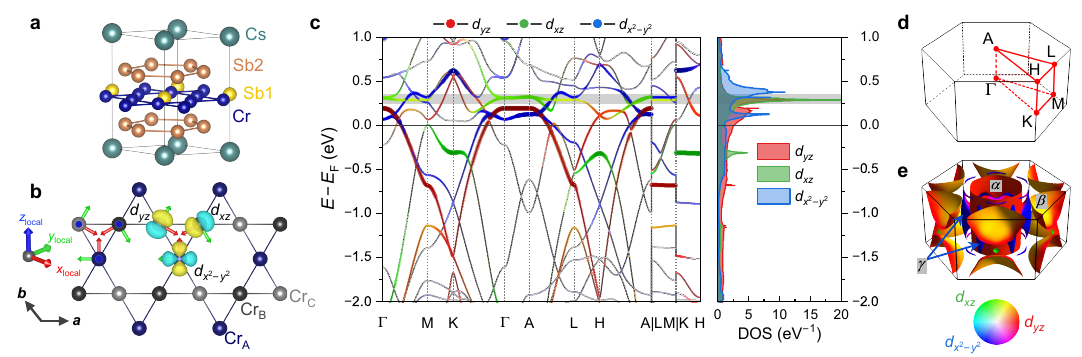}
  \caption{\textbf{Crystal and electronic structures for \Cr135 at 5 GPa.} \textbf{a} Crystal structure for \Cr135. \textbf{b} Top view of the kagome Cr-lattice. The local coordinates and atomic $d$ orbitals are also shown for different Cr-sublattices. \textbf{c} Band structures and projected density of states for \Cr135. \textbf{d} The first Brillouin zone and high-symmetry lines for hexagonal \Cr135 lattice. \textbf{e} Fermi surfaces for \Cr135, plotted with FermiSurfer program \cite{Kawamura-2019-Fermisurfer}. The orbital contributions of Cr-$d_{yz}$, Cr-$d_{xz}$, and Cr-$d_{x^2-y^2}$ in \textbf{c} and \textbf{e} are denoted by red, green, and blue color components, respectively. The crystal structures are constructed with VESTA program \cite{Momma-2011-Vesta}. }
  \label{fig:1}
\end{figure*}

\textbf{First-principles calculations.} We first conducted first-principles structural optimizations on \Cr135 under external pressure of 5 GPa. The relaxation yields a hexagonal lattice with parameters of $a =$ 5.272 \AA, $c =$ 8.400 \AA. The band structures and projected density of states for the optimized structure are shown in Fig.~\ref{fig:1}c. Evidently, most bands near \Ef\ are mainly contributed by the Cr-$d_{xz}$, Cr-$d_{yz}$, and Cr-$d_{x^2-y^2}$ orbitals. As shown in Fig.~\ref{fig:1}e, the $d_{yz}$ orbitals form a quasi-two-dimensional (Q2D) cylindrical Fermi surface around the $\Gamma$ point (namely, the $\alpha$ band); while around the L point, the $d_{xz}$ orbitals strongly hybridize with the $d_{yz}$ orbitals and form six three dimensional (3D) FS pockets ($\beta$ band). Surrounding the Q2D cylindrical FS, the $d_{x^2-y^2}$ orbitals form another set of six separated ridge-shaped FS sheets ($\gamma$ band). It is notable that at $E - $\Ef\ $ \sim $ 0.3 eV, the $d_{xz}$ orbitals hybridize with $d_{yz}$ and form an extremely flat band within the $k_y=0$ plane (as illustrated by $\Gamma$-M-L-A segments). Such a flat band is responsible for the highest DOS peak shown in the right hand side of Fig.~\ref{fig:1}c. It can also be shown to arise from the frustrated geometry of the kagome Cr lattice \cite{SI,Jiang-2023-FeGe}. In fact, two sets of moderately distorted kagome bands could be recognized from the \Cr135 band structures near \Ef\ \cite{SI}, with filling levels slightly below the IFBs. We will show later that the interband processes between the IFBs and the highest dispersive band (HDB) substantially contribute, and more interestingly, determine the momentum dependence of spin fluctuations and SC orders in \Cr135.

\textbf{RPA analyses.} To explore the instabilities in the pressurized \Cr135, we conducted RPA analyses on the DFT calculated electronic structures. Our calculations stem from a model Hamiltonian of $H = H_0 + \sum_{\tau} H^{\tau}_{int}$. Here $H_0$ is the non-interacting Hamiltonian obtained from Wannier downfolding, for which 30 locally defined Cr-3$d$ and Sb-5$p$ atomic orbitals are adopted as initial guess. The interacting $H^{\tau}_{int}$ is considered for the Cr sites, and is given by:
\begin{equation}
  H^{\tau}_{int} = U \sum_{i} n_{i \uparrow} n_{i \downarrow}+U^{\prime} \sum_{i<j} n_{i} n_{j} +J \sum_{i<j, \sigma, \sigma^{\prime}} c_{i \sigma}^{\dagger} c_{j \sigma^{\prime}}^{\dagger} c_{i \sigma^{\prime}} c_{j \sigma}+J^{\prime} \sum_{i \neq j} c_{i \uparrow}^{\dagger} c_{i \downarrow}^{\dagger} c_{j \downarrow} c_{ j \uparrow},
\end{equation}
where $i$, $j$ are the $d$-orbital indices within site $\tau$.
It is worth noting that all the obtained Cr-$d$ Wannier orbitals are well-localized (with spreading of only $\sim$ 1.5 \AA$^2$) and their atomic symmetries are well maintained. Therefore we can adopt symmetry relations that $J = J^{\prime}$, $U = U^{\prime} + 2J$, which is also consistent with our constrained RPA results.
The bare electron, spin, and charge susceptibilities are then calculated using:
\begin{subequations}\label{chiRPA}
  \begin{gather}
     {\chi_0}_{st}^{\space pq}\left(\mathbf{Q}, i \nu_m\right)=\quad-\frac{1}{N \beta} \sum_{\mathbf{k}, i \omega_n} G_{sp}\left(\mathbf{k}, i \omega_n\right) G_{qt} \left(\mathbf{k}+\mathbf{Q}, i \omega_n+i \nu_m\right), \label{chi0} \\
    \chi_\mathrm{S}=\chi_0\left(1-U_\mathrm{S} \chi_0\right)^{-1}, \label{chiS} \\
    \chi_\mathrm{C}=\chi_0\left(1+U_\mathrm{C} \chi_0\right)^{-1}. \label{chiC}
  \end{gather}
\end{subequations}
Here we have restricted our calculations to the real part of the $i\nu_m = 0$ term \cite{Graser-2009-FeSC}. We shall focus on the property of the susceptibility matrix block $\chi^{dd}$ that contracts with the Coulomb interaction vertices (i.e., $pq$, $st \in$ nonzero $U$ matrix indices), since they will dominate the spin and charge susceptibilities near the critical point.

\begin{figure}
  \includegraphics[width = 9 cm]{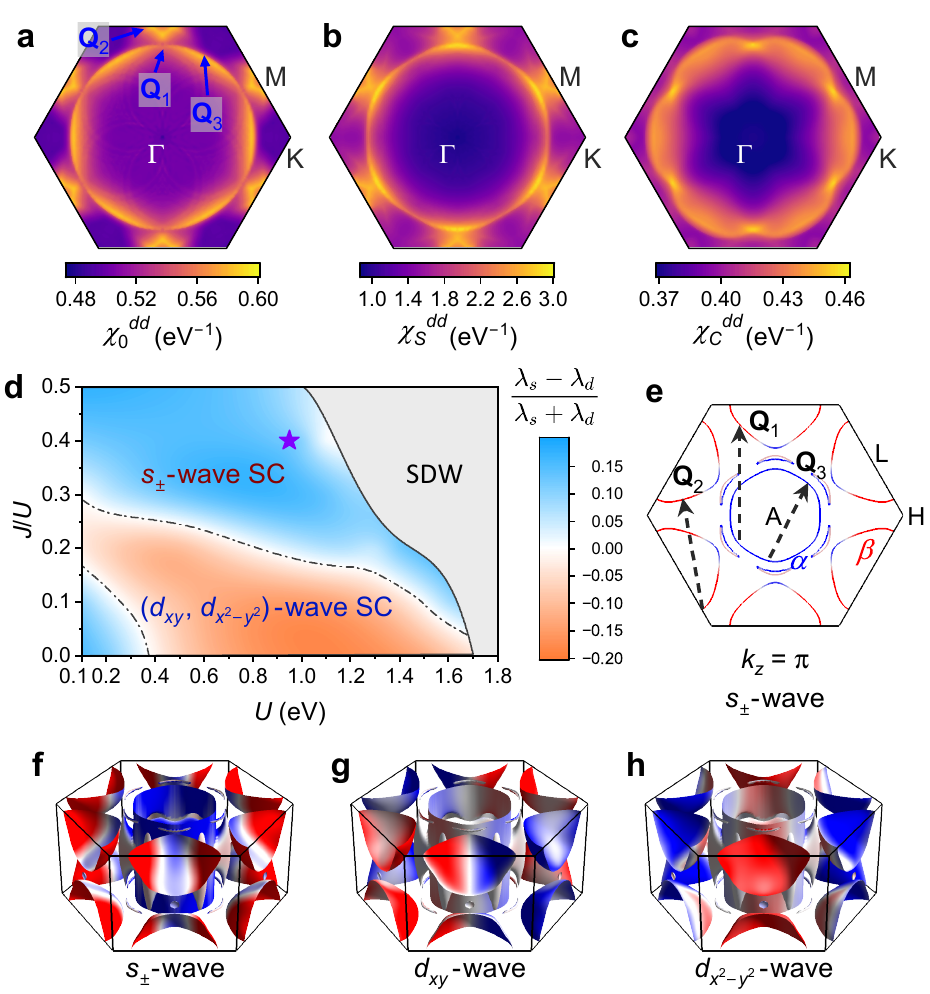}
  \caption{\textbf{Results for RPA analyses.} \textbf{a-c} The largest eigenvalues for calculated bare electron ($\chi_0^{dd}$), spin ($\chi_\mathrm{S}^{dd}$), and charge ($\chi_\mathrm{C}^{dd}$) susceptibilities, under $U = 0.95$ eV, $J/U = 0.4$, $1/\beta = 1$ meV. \textbf{d} The RPA phase diagram in terms of the leading instabilities. The purple star denotes the interactions estimated from constrained RPA method (i.e., $U = 0.95$ eV, $J/U = 0.4$). \textbf{e} Nesting vectors of the Fermi surface, plotted with superconducting gap of the $s_\pm$ phase. \textbf{f-h} Superconducting gap functions for the leading solutions.}
  \label{fig:2}
\end{figure}

Figs.~\ref{fig:2}a-c show the calculation results under $U = 0.95$ eV, $J/U = 0.4$, at $\beta = $ 1000 eV$^{-1}$. In Fig.~\ref{fig:2}a, a set of V-shaped peaks around the M points and a set of ring-shaped peaks surrounding the interior of the first Brillouin zone could be distinguished for the leading eigenvalue of $\chi_0^{dd}$. Among them, three peaks at $\mathbf{Q}_1 = $(0.0, 0.41, 0.0), $\mathbf{Q}_2 = $($-$0.06, 0.5, 0.0), and $\mathbf{Q}_3 = $(0.2, 0.34, 0.0) are intimately connected to the FS nesting. As shown in Fig.~\ref{fig:2}e, $\mathbf{Q}_1$, $\mathbf{Q}_2$, and $\mathbf{Q}_3$ correspond to the inter-pocket scattering between $\alpha$ and $\beta$ bands, inter-pocket scattering between different $\beta$ band FS pockets, and intra-pocket scattering within $\alpha$ band, respectively. As the interactions are switched on, the peaks are generally enhanced in $\chi_\mathrm{S}^{dd}$ and suppressed in $\chi_\mathrm{C}^{dd}$, indicating the dominance of the spin fluctuations over charge ones. Specifically, the peaks at $\mathbf{Q}_1$ and $\mathbf{Q}_2$ exhibit more pronounced amplifications than $\mathbf{Q}_3$. In the context of spin-fluctuation mediated SC paring, the strong enhancement at $\mathbf{Q}_1$ and $\mathbf{Q}_2$ implies strong repulsive pairing interactions between different FS pockets. Note that the pair scatterings of $\mathbf{Q}_1$ favor a sign change between $\alpha$ and $\beta$ bands, while the scatterings of $\mathbf{Q}_2$ favor a sign change between different FS pockets of $\beta$ band. The interplay of various scattering processes then indicates a possible competitions between different SC orders in \Cr135 \cite{Kuroki-2008-LaFeAsO,Graser-2009-FeSC}.

The spin-fluctuation mediated SC gap symmetries are then investigated by solving the linearized gap equations. As shown in Fig.~\ref{fig:2}d, under moderate $U$ (0.5 $\sim$ 1.2 eV), the leading gap symmetry is mainly tuned by the ratio of $J/H$. When $J/H$ is large, the system tends to form an SC gap with $A_{1g}$ symmetry [i.e., the $s_{\pm}$-wave solution shown in Fig.~\ref{fig:2}f]. When $J/H$ is small, on the other hand, the SC gap with $E_{2g}$ symmetry [the ($d_{xy}$, $d_{x^2-y^2}$)-wave in Fig.~\ref{fig:2}g,h] takes place as the leading solution. To gain insight into the SC gap symmetry in the experimental system, we performed constrained RPA (cRPA) calculations to estimate the interaction parameters in the pressurized \Cr135. The average interaction parameters are $U = 0.95$ eV and $J/U = 0.4$, indicating that the system resides within the $s_{\pm}$-wave region. Fig.~\ref{fig:2}f illustrates the FS distribution of the gap function for the $s_{\pm}$-wave solution. There is no sign change on the $\alpha$-pocket, consistent with common understanding of $s$-wave. However, within each $\beta$ FS patch, it reveals an intriguing sign-changing character. In this $s_{\pm}$-wave SC phase, line nodes exists without crystal symmetry breaking. Such a feature could be detected through the deviation of full-gap behavior in heat capacity and penetration depth measurements \cite{Pang-2015-K233}. Furthermore, for the competing $d$-wave solution, there is a degree of freedom in forming linear combinations of the $d_{xy}$ and $d_{x^2-y^2}$ bases. This can result in nematic $d_{xy}$ and $d_{x^2-y^2}$ phases that break the rotational symmetry, or the chiral $d \pm id$ phases that break the time reversal symmetry (TRS) \cite{Kiesel-2012-MeanField,Kiesel-2012-sublattice}. The symmetry of SC order in \Cr135\ could be further identified via spatial anisotropic experiments such as angle-resolved critical field measurements, as well as TRS sensitive experiments such as muon spin rotation and polar Kerr effect studies \cite{Venderbos-2016-Hc2,Kallin-2016-Chiral}.

\begin{figure}
  \includegraphics[width = 10 cm]{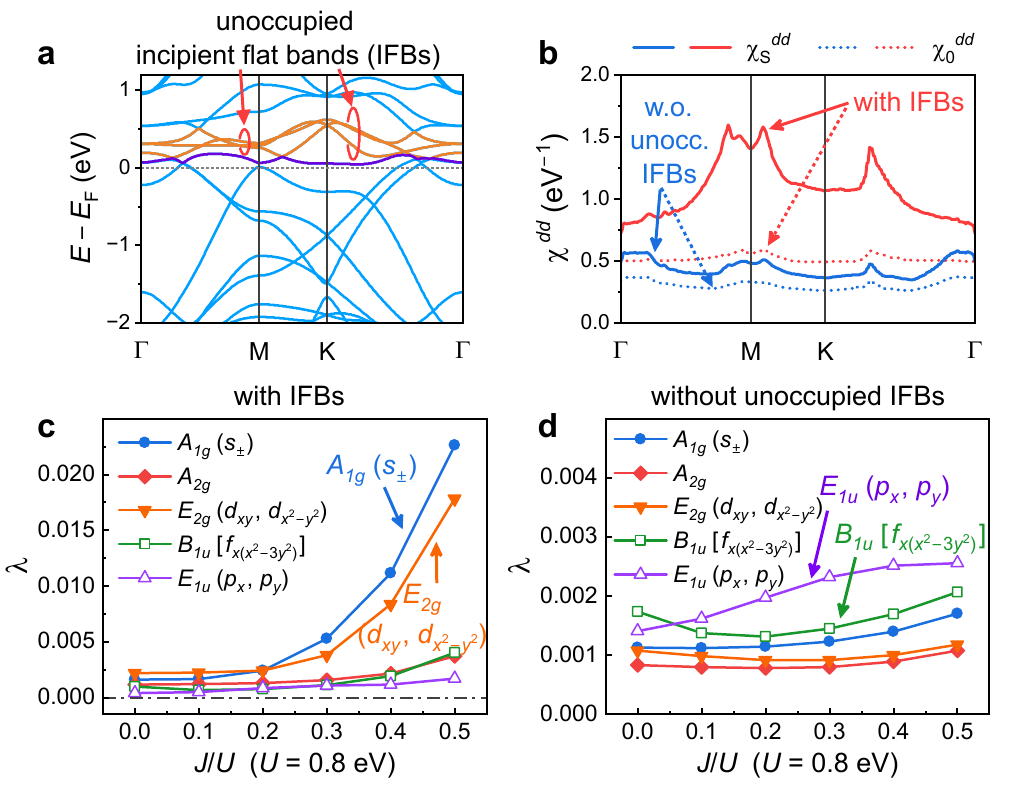}
  \caption{\textbf{Momentum selective contributions to the spin fluctuations from the unoccupied incipient flat bands (IFBs).} \textbf{a} Band structure for \Cr135. The unoccupied and partially occupied incipient flat bands are highlighted by orange and purple lines, respectively. \textbf{b} The largest eigenvalues for spin and bare electron susceptibility with and without unoccupied IFBs. \textbf{c,d} Leading eigenvalues of the linearized gap equations calculated with and without unoccupied IFBs, respectively.}
  \label{fig:3}
\end{figure}

\textbf{Role of incipient flat bands.} From the result of our first-principles calculations, we have manifested the presence of IFBs in \Cr135 (as highlighted in Fig.~\ref{fig:3}a). Given the fact that these bands contribute to prominent DOS peaks around \Ef, it is therefore natural to explore their role in the aforementioned scenario of superconductivity in \Cr135. In fact, the significance of the IFBs has been usually investigated as incipient flat parts of a dispersive band \cite{Linscheid-2016-FeSe,Yamazaki-2020-BCO}. Here we emphasize that, even if the full part of the IFBs are flat and unoccupied, they still substantially affect the momentum dependence of spin fluctuations and thus are critical to the SC orders in \Cr135.

To elucidate the crucial role of unoccupied IFBs, we conducted calculations both with and without the unoccupied IFBs for comparison. As shown in Fig.~\ref{fig:3}b, strikingly, the AFM peaks of spin susceptibility $\chi_S^{dd}$ are significantly suppressed after we exclude the unoccupied IFBs, leaving a rather broad hump around the $\Gamma$ point. The result clearly demonstrates that the AFM fluctuations are selectively and substantially contributed by the unoccupied IFBs. As a consequence, the spin-fluctuation mediated SC orders are also significantly altered. Figs.~\ref{fig:3}c,d depict the leading solutions of linearized gap equations calculated with and without unoccupied IFBs, respectively. It is evident that without the unoccupied IFBs, the AFM spin-fluctuation mediated solutions (e.g., the $s_{\pm}$-wave and the $d$-wave solutions) are remarkably suppressed. In contrast, owing to the surviving FM fluctuations, two spin-triplet solutions of $E_{1u}$ ($p$-wave) and $B_{1u}$ ($f$-wave) symmetries are stabilized and turn to be the leading SC orders.

\begin{figure}
  \includegraphics[width = 16 cm]{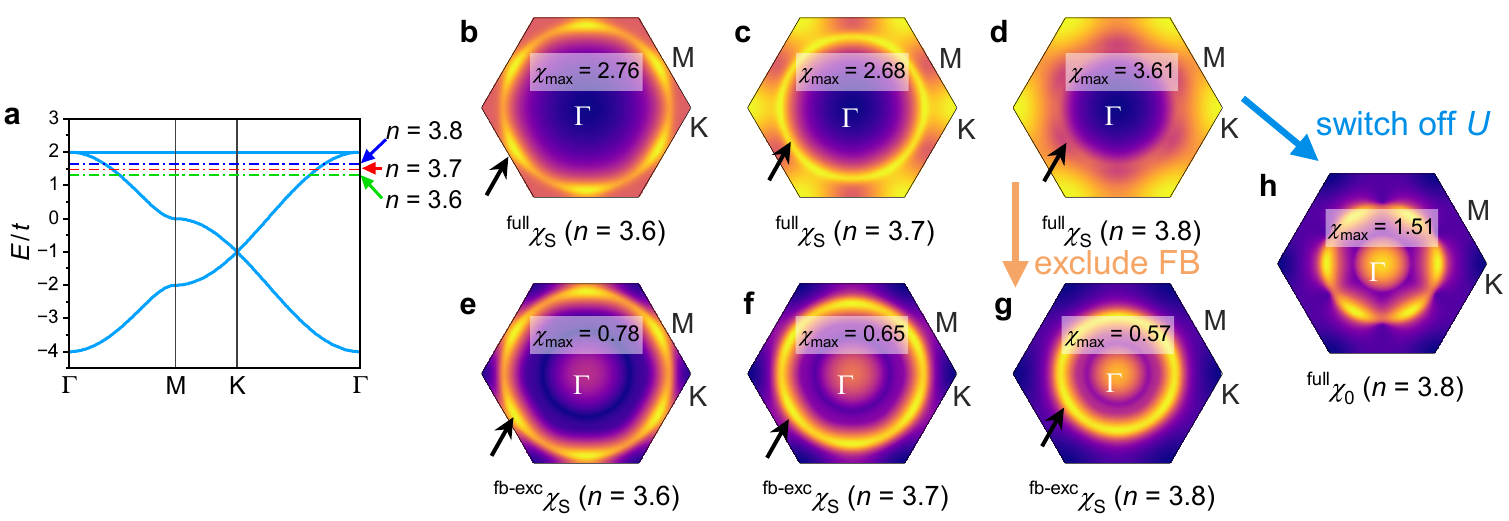}
  \caption{\textbf{Flat-band induced enhancement of the AFM spin fluctuations in kagome Hubbard model.} \textbf{a} Band structure of the ideal kagome model. The green, blue, and red dash-dot lines correspond to the Fermi level at filling level of $n =$ 3.6, 3.7, and 3.8, respectively. \textbf{b-g} Leading eigenvalues of spin susceptibilities for kagome Hubbard model under different filling levels, calculated at $U = 2t$. \textbf{b-d} are calculated with all three bands, \textbf{e-g} are calculated without the flat band. \textbf{h} Leading eigenvalues of the bare electron susceptibility at filling level of $n =$ 3.8. The black arrows point to the ring-shaped peak sets originating from the Fermi surface nesting, which do not vary upon excluding the flat band, while the AFM peaks around the K points are intimately related to the flat band.}
  \label{fig:4}
\end{figure}

The IFB induced enhancement of the AFM spin fluctuations is a ubiquitous feature for Kagome systems where the \Ef\ resides between the flat bands and the dispersive bands.
As shown in Fig.~\ref{fig:4}b, for kagome Hubbard model at filling level of $n =$ 3.6 (i.e., 1.2 electron per site), the spin susceptibility $\chi_\mathrm{S}$ exhibits a ring-shaped peak set which originates from the two-dimensional cylindrical FS. When we move \Ef\ closer to the flat band by upshifting the filling level to $n=$ 3.7 and 3.8 (Fig.~\ref{fig:4}c,d), the size of the ring gradually shrinks as the FS shrinks towards the $\Gamma$ point. Even more prominently, another set of AFM peaks around the K points emerge and quickly become dominant, thus the ring-shaped structure is difficult to be distinguished from the background in Fig. ~\ref{fig:4}d. The ring-shaped structure can be clearly identified in calculations excluding the contributions from the flat bands (Figs.~\ref{fig:4}e-g) or by switching off the Hubbard $U$ terms (fig.~\ref{fig:4}h). In either cases, the AFM peaks around the K points disappear. Therefore the enhancement of AFM fluctuations should be a consequence of the combination of both the flat band and the local Coulomb interaction.

\section*{Discussion}

To comprehend the embedding mechanism behind the IFB-induced momentum-dependent enhancement of spin fluctuations, we rewrite Eq.~\eqref{chi0} into the spectral form that:
\begin{equation} \label{chi0-spec}
    {\chi_0}_{st}^{pq}(\mathbf{Q})=-\frac{1}{N} \sum_{\mathbf{k}, \mu \nu} \frac{f\left(E_{\mathbf{k}+\mathbf{Q}, \nu}\right)-f\left(E_{\mathbf{k}, \mu}\right)}{E_{\mathbf{k}+\mathbf{Q}, \nu}-E_{\mathbf{k}, \mu}+\mathrm{i} 0^{+}} \times \braket{s|\mathbf{k}, \mu} \braket{\mathbf{k}, \mu|p} \braket{q|\mathbf{k}+\mathbf{Q}, \nu} \braket{\mathbf{k}+\mathbf{Q}, \nu|t},
\end{equation}
where $f(E)$ is the Fermi-Dirac distribution function, $\mu$ and $\nu$ are band indices. For the channels involving unoccupied IFBs (without loss of generality, here we set $\nu$ to be the band index for unoccupied IFBs), $f\left(E_{\mathbf{k}+\mathbf{Q}, \nu}\right)$ is fixed to be 0 and $E_{\mathbf{k}+\mathbf{Q}, \nu}$ is nearly a constant. Therefore the Lindhard function [i.e., the $\frac{f\left(E_{\mathbf{k}+\mathbf{Q}, \nu}\right)-f\left(E_{\mathbf{k}, \mu}\right)}{E_{\mathbf{k}+\mathbf{Q}, \nu}-E_{\mathbf{k}, \mu}+\mathrm{i} 0^{+}}$ part] has negligible momentum dependence and the AFM feature should mainly emerge from the orbital or site degrees of freedom.

\begin{figure}
  \includegraphics[width = 10 cm]{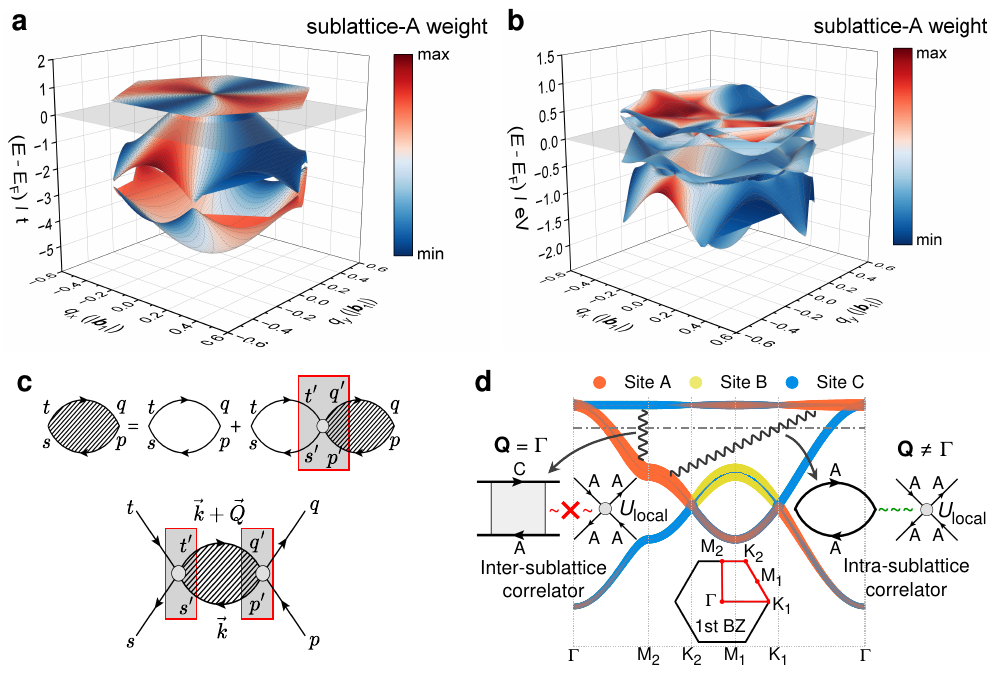}
  \caption{\textbf{Sublattice-momentum coupling as a driving mechanism for the AFM fluctuations.} \textbf{a,b} Band structures and sublattice projections for Cr$_{\mathrm{A}}$ in ideal kagome model and downfolded $M_z$-odd Wannier Hamiltonian of \Cr135, respectively. \textbf{c} Feynman diagrams for the Dyson series and the pairing vertex of RPA calculations. The locality of the Coulomb interactions is emphasized by the red rectangles. \textbf{d} Schematic illustration for the SMC-driven mechanism of the AFM enhancement in the ideal kagome model: The dominant inter-sublattice correlator at $Q = \Gamma$ does not appear in the Wick contractions of the local Coulomb interactions, while the intra-sublattice correlator at $Q \neq \Gamma$ does.}
  \label{fig:5}
\end{figure}

Figs.~\ref{fig:5}a,b show the sublattice character for the ideal kagome bands and \Cr135 mirror-odd downfolded bands (which is responsible for the leading fluctuation channel in \Cr135, see supplemental materials for detail \cite{SI}), respectively. Evidently, the sublattice degree of freedom is strongly coupled to the momentum and exhibits similar SMC pattern for both ideal kagome model and \Cr135. For the flat band, the sublattice-A weight is mainly distributed around the $\Gamma$-K$_1$ line; while for the HDB, the sublattice-A weight peaks around the M$_2$ point. When the Fermi level cuts the top of the HDB and makes the flat band incipient, the SMC pattern will be encoded in the inter-band processes and become crucial for the spin fluctuations of the system.

To see how the SMC feature affects the spin susceptibilities in the presence of local Coulomb interactions, let's consider two processes depicted by the Feynman diagrams in Fig.~\ref{fig:5}c. As highlighted by the red rectangles in Fig.~\ref{fig:5}c, the indices of paramagnons [i.e., ($p^{\prime}$, $q^{\prime}$) or ($s^{\prime}$, $t^{\prime}$) in Fig.~\ref{fig:5}c] are restricted to the same site by the locality of Coulomb interactions. The restriction requires that the propagators of both $\mathbf{k}$ and $\mathbf{k}+\mathbf{Q}$ must have the same sublattice component. Therefore, for the Dyson series shown in the upper diagram of Fig.~\ref{fig:5}c, the correlators for \emph{intra-sublattice} paramagnons are selectively enhanced. And for the SC pairing vertex shown in the lower diagram of Fig.~\ref{fig:5}c, \textit{only intra-sublattice} paramagnon correlators are picked out to contribute the SC pairing. As illustrated in Fig.~\ref{fig:5}d, in kagome lattice, the sublattice weight separation in occupied (dispersive) and unoccupied (flat) bands makes the paramagnons of $\mathbf{Q} = \Gamma$ \emph{inter-sublattice}. Thus only \emph{intra-sublattice} channels at $\mathbf{Q} \neq \Gamma$ are involved in the diagrammatic summation of Fig.~\ref{fig:5}c. As a consequence, the AFM fluctuations are selectively enhanced by the local Coulomb interactions.

The SMC is a general feature for the line graphs of planar bipartite lattices \cite{Mielke-1992-FM}. In fact, similar k-space sublattice weighting could also be observed for other kagome systems such as CsV$_3$Sb$_5$ and CoSn \cite{Wu-2021-V135,Kang-2020-CoSn}. In CsV$_3$Sb$_5$, the SMC manifests as a set of $p$-type vHSs (where the saddle point states correspond one-to-one with the sublattice sites), alongside a set of $m$-type vHSs (where two of the three sublattice components mix at each saddle point). The interplay between the SMC and FS nesting in CsV$_3$Sb$_5$ yields a unique sublattice interference character which selectively shapes the contributions of different Coulomb interaction components \cite{Park-2021-vHS,Wu-2021-V135}. In \Cr135, on the other hand, the kagome IFBs have a more pronounced influence on the system's low energy physics. Due to their flatness and unoccupied nature, these IFBs lead to a stronger coupling of the SMC with the locality of Coulomb interactions, which results in a selective impact on the momentum of the leading fluctuation channels. In addition, due to the self-doping effect between the Sb1-$p_z$ orbitals and the Cr-$d$ orbitals, the position of IFBs in \Cr135 is relatively inertia to external pressure, and thus the SMC mechanism is expected to be present in this compound within a wide range of pressure (see supplemental materials for details \cite{SI}). The SMC in these systems establishes a connection between the reciprocal space band structures and the real-space patterns of Coulomb interactions and spin/charge orders. We anticipate that it can be further utilized to explore a wider variety of novel physical phenomena in geometrically frustrated systems.

In conclusion, we have performed first-principles calculations and RPA investigations on pressurized \Cr135, at external pressure of 5 GPa. Our calculations reveal strong AFM fluctuations in \Cr135, which mediate a dominating $s_{\pm}$-wave and a competing ($d_{xy}$, $d_{x^2-y^2}$)-wave SC order. Under the dominance of local Coulomb interactions, the k-space sublattice weight separation of occupied bands and unoccupied IFBs inhibits the formation of intra-sublattice FM fluctuations and thus selectively enhances the AFM fluctuations. Such SMC-enhanced AFM mechanism is expected to be widely exist in kagome systems where the Fermi level resides between IFBs and dispersive bands, and may also apply to other geometrically frustrated systems.

\section*{Methods}
\subsection{First-principles calculations}
The first-principles calculations are performed with density functional theory (DFT), as implemented in the Vienna Ab initio Simulation Package (VASP) \cite{Kresse-1993-VASP}. The Kohn-Sham wave functions are represented under projected augmented wave (PAW) basis \cite{Blochl-1994-PAW}. The crystal structure of \Cr135 is optimized at external pressure of 5 GPa, with a solid revised Perdew-Burke-Ernzerhof (PBEsol) exchange correlation functional \cite{Perdew-2008-PBEsol}. For the self-consistent and density of state (DOS) calculations, we employed a 12$\times$12$\times$8 and a 24$\times$24$\times$12 $\Gamma$-centered $k$ mesh respectively. Our tight-binding Hamiltonian is then constructed via Wannier downfolding \cite{Mostofi-2008-Wannier90}, where 30 atomic like projectors of 15 Cr-$d$ and 15 Sb-$p$ are adopted as initial seeds. The local axes are set that the three Cr sites are connected by successive $C_3$ rotations, as shown in Fig.~1b. Details of band structure downfolding are given in the supplemental materials \cite{SI}.

\subsection{Random phase approximation analysis}

Our RPA calculations begin from the model Hamiltonian that:
\begin{equation}
H = H_0 + \sum_{\tau} \left[ U \sum_{i} n_{\tau, i \uparrow} n_{\tau, i \downarrow}+U^{\prime} \sum_{i<j} n_{\tau, i} n_{\tau, j} + J \sum_{i<j, \sigma, \sigma^{\prime}} c_{\tau, i \sigma}^{\dagger} c_{\tau, j \sigma^{\prime}}^{\dagger} c_{\tau, i \sigma^{\prime}} c_{\tau, j \sigma}+J^{\prime} \sum_{i \neq j} c_{\tau, i \uparrow}^{\dagger} c_{\tau, i \downarrow}^{\dagger} c_{\tau, j \downarrow} c_{\tau, j \uparrow} \right],
\end{equation}
where $H_0$ is the bare-election tight-binding Hamiltonian obtained from Wannier downfolding; $\tau$ is the Cr site index, $i$, $j$ are $d$ orbital indices for each Cr. The Coulomb parameters follow the general symmetry relation that $J = J^{\prime}$, $U = U^{\prime} + 2J$.

The bare electron susceptibility is calculated using:
\begin{equation} \label{chi0_method}
{\chi_0}^{pq}_{st}(\mathbf{q}, \mathrm{i}\nu_m)=-\frac{1}{N_{\mathbf{k}}\beta}\sum_{n,\mathbf{k}}G_{sp}(\mathrm{i}\omega_n, \mathbf{k})G_{qt}(\mathrm{i}\omega_n+\mathrm{i}\nu_m, \mathbf{k+q}),
\end{equation}
where $\mathrm{i}\omega_n$ and $\mathrm{i}\nu_m$ are fermionic and bosonic Matsubara frequencies respectively; $p$, $q$, $s$, $t$ are orbital indices; $G_{sp}(\mathrm{i}\omega_n, \mathbf{k}) = \sum_{\mu} \frac{\braket{s|\mathbf{k},\mu}\braket{\mathbf{k},\mu|p}}{\mathrm{i}\omega_n - E_{\mathbf{k}, \mu}}$ is the bare electron propagator. We have adopted a 100$\times$100$\times$30 $Q$ mesh and $\frac{1}{\beta} =$ 0.001 eV for the calculations.

For spin and charge susceptibilities, we employ matrix random-phase approximation (RPA) formula that:
\begin{subequations}
  \begin{align}
    &\chi_\mathrm{S}=\chi_0\left(1-U_\mathrm{S} \chi_0\right)^{-1}, \\
    &\chi_\mathrm{C}=\chi_0\left(1+U_\mathrm{C} \chi_0\right)^{-1}.
  \end{align}
\end{subequations}
The matrix elements for spin and charge vertices are:
\begin{equation}
{U_\mathrm{S}}^{pq}_{st}=
      \begin{cases}
        U,          & p=q=s=t, \\
        J,          & p=q \neq s=t, \\
        J,          & p=t \neq q=s, \\
        U^{\prime}, & p=s \neq q=t,
      \end{cases}
\quad
{U_\mathrm{C}}^{pq}_{st}=
      \begin{cases}
        U,               & p=q=s=t, \\
        2U^{\prime} - J, & p=q \neq s=t, \\
        J,               & p=t \neq q=s, \\
        2J - U^{\prime}, & p=s \neq q=t.
      \end{cases}
\end{equation}
The superconducting (SC) pairing vertex functions for spin-singlet and spin-triplet channels are then calculated by:
\begin{subequations}
    \begin{align}
       & V^{\mathrm{singlet}}(\mathbf{k}^{\prime}q,-\mathbf{k}^{\prime}s;\mathbf{k}p,-\mathbf{k}t) = \frac{1}{4} (3U_{\mathrm{S}} + U_{\mathrm{C}})^{pq}_{st} + \Bigl\{ \frac{1}{4} {\left[ 3U_{\mathrm{S}}\chi_{\mathrm{S}}(\mathbf{k}^{\prime}-\mathbf{k})U_{\mathrm{S}} - U_{\mathrm{C}}\chi_{\mathrm{C}}(\mathbf{k}^{\prime}-\mathbf{k})U_{\mathrm{C}}\right]}^{pq}_{st} + (\mathbf{k}p \leftrightarrow -\mathbf{k}t) \Bigr\}, \\
       & V^{\mathrm{triplet}}(\mathbf{k}^{\prime}q,-\mathbf{k}^{\prime}s;\mathbf{k}p,-\mathbf{k}t) = \frac{1}{4} (-U_{\mathrm{S}} + U_{\mathrm{C}})^{pq}_{st} - \Bigl\{ \frac{1}{4} {\left[ U_{\mathrm{S}}\chi_{\mathrm{S}}(\mathbf{k}^{\prime}-\mathbf{k})U_{\mathrm{S}} + U_{\mathrm{C}}\chi_{\mathrm{C}}(\mathbf{k}^{\prime}-\mathbf{k})U_{\mathrm{C}}\right]}^{pq}_{st} - (\mathbf{k}p \leftrightarrow -\mathbf{k}t) \Bigr\}. \label{V_SC_T}
    \end{align}
\end{subequations}
Here we only take the zero-frequency component of $\chi$. Note that in the SC pairing vertex calculations, only $pq$, $st$ $\in$ nonzero $U$ matrix indices contribute to the result. Thus we can restrict our calculations to the matrix block of these indices. Here we denote the matrix block by $\chi^{dd}$. It is straightforward to see that we can simplify the calculations by replacing all $\chi_0, \chi_\mathrm{S}$, and $\chi_\mathrm{C}$ matrices with only $dd$ block in \eqref{chi0_method}-\eqref{V_SC_T} without any loss of the SC pairing vertex Information.

The weak-coupling SC gap equation is then constructed as \cite{Graser-2009-FeSC}:
\begin{equation} \label{V_SC_fs_1}
    \lambda \Delta(\mathbf{k}^{\prime}) = - \frac{1}{V_\mathrm{BZ}} \int_{FS} \frac{\mathrm{d}^2k_{\parallel}}{|v_{\mathbf{k}}^{\perp}|} V(\mathbf{k}^{\prime},\mathbf{k}) \Delta(\mathbf{k}),
\end{equation}
where the integration is done over the Fermi surface (FS), $v_{\mathbf{k}}^{\perp}$ corresponds to the Fermi velocity of the FS $\mathbf{k}$ points.

\section*{Data availability statement}
The authors declare that the data supporting the findings of this study are available within the paper and its supplementary information files.

\section*{Code availability statement}
The codes used in this study will be available upon reasonable request.

%

\section*{Acknowledgements}
The authors would like to thank Xi Dai, Jianhui Dai, Lunhui Hu and Yi Liu for fruitful discussions. 
This work was supported by NSFC (12350710785, 12274364, 12304175, and 11874137), the National Key R\&D Program of China (Nos. 2022YFA1402202, 2024YFA1408303, 2023YFA1406101, and 2022YFA1403202), and the Key R\&D Program of Zhejiang Province (2021C01002).
The calculations were performed on clusters at the Center of Correlated Matters Zhejiang University and the High Performance Computing Center at Hangzhou Normal University.

\section*{Author contributions}
C. Cao and G.-H. Cao proposed the original idea. S. Wu wrote the RPA code and performed the calculations. C. Cao and C. Xu provided technical assistance in computing and code debugging. S. Wu and C. Cao wrote the manuscript. All authors discussed the results and commented on the manuscript.

\section*{Competing interests}
The authors declare no competing interests.

\newpage 

\end{document}